\documentclass[%
 aip,
 amsmath,amssymb,
 reprint,%
]{revtex4-1}

\usepackage[pdftex]{graphicx}
\usepackage{dcolumn}
\usepackage{bm}

\usepackage[utf8]{inputenc}
\usepackage[T1]{fontenc}
\usepackage{mathptmx}
\usepackage{etoolbox}

\makeatletter
\def\@email#1#2{%
 \endgroup
 \patchcmd{\titleblock@produce}
  {\frontmatter@RRAPformat}
  {\frontmatter@RRAPformat{\produce@RRAP{*#1\href{mailto:#2}{#2}}}\frontmatter@RRAPformat}
  {}{}
}%
\makeatother
\begin{document}

\preprint{AIP/123-QED}

\title{Augmentation of the Electron Counting Rule with Ising Model}

\author{Karol Kawka}
\author{Paweł Kempisty}%
\author{Konrad Sakowski}%
 \altaffiliation[Also at ]{Institute of Applied Mathematics and Mechanics, University of Warsaw, Poland.}
\author{Stanisław Krukowski}%
\author{Michał Boćkowski}%
\affiliation{%
Institute of High Pressure Physics, Polish Academy of Sciences, Sokolowska 29/37, 01-142 Warsaw, Poland
}%

\author{\\David Bowler}
\affiliation{
 London Centre for Nanotechnology, UCL, 17-19 Gordon St, London, WC1H 0AH, UK
}

\author{Akira Kusaba}
 \email{kusaba@riam.kyushu-u.ac.jp}
\affiliation{%
 Research Institute for Applied Mechanics, Kyushu University, Fukuoka 816-8580, Japan
}%

\date{\today}

\begin{abstract}
On semiconductor growth surfaces, surface reconstructions appear. Estimation of the reconstructed structures is essential for understanding and controlling growth phenomena.
In this study, the stability of a mixture of two different surface reconstructions is investigated. Since the number of candidate structures is enormous, the structures sampled by Bayesian optimization are analyzed. As a result, the local electron counting (EC) rule alone was found to be insufficient to explain such stability. Then, augmenting the EC rule, a data-driven Ising model is proposed. The model allows the evaluation of the whole enormous number of candidate structures. The approach is expected to be useful for theoretical studies of such mixtures on various semiconductor surfaces. 
\end{abstract}

\maketitle

\section{INTRODUCTION}
Semiconductor surfaces with unstable dangling bonds are generally stabilized by dimer formation or chemisorption, that is, surface reconstructions, which change adsorption/desorption energy and surface migration potential in addition to surface energy.
So, estimating the realistic surface structure during epitaxial growth is essential for correct understanding and quantitative modeling of growth phenomena \cite{bui2023insight,boero2022atomistic,bui2019first,bui2018reaction}. The surface reconstruction in vapor phase epitaxy under atmospheric pressure has been studied primarily on the basis of first-principles calculations since an in-situ observation with electron beam is difficult in such an environment \cite{fritsch1998ab,pignedoli2001dissociative,van2002first,northrup2004indium,northrup2004strong,suzuki2007theoretical,akiyama2010surface,akiyama2012ab,dreyer2014absolute}. Incorporating the entropy of gas molecule based on statistical thermodynamics into the first-principles energetics, theoretical surface phase diagrams have been created that map an experimental condition (i.e., temperature and partial pressures) to the thermodynamically most stable surface reconstruction \cite{kangawa2001new,van2002first,kangawa2002theoretical,northrup2004indium,northrup2004strong,ito2008ab,akiyama2010surface,akiyama2012ab,kangawa2013surface,dreyer2014absolute,kempisty2019evolution}. 

GaN is a promising semiconductor material for next-generation power devices. In GaN(0001) metalorganic vapor phase epitaxy (MOVPE), which is the industrially most standard growth system, typical experimental conditions are mapped near the phase boundary between the Ga-adsorbed surface, Ga$_{ad}$(2$\times$2) \cite{smith1997reconstructions,rapcewicz1997theory}, and H-adsorbed surface, 3Ga-H(2$\times$2) \cite{fritsch1998ab,kusaba2017thermodynamic}. That is, one is the most stable structure and the other is the metastable structure. Note that such a theoretical result is based on discussions only within a small size of lateral cell, i.e., GaN(0001)-(2$\times$2). 
It is also known that NH$_3$/NH$_x$/H-adsorbed surfaces appear depending on partial pressures \cite{kempisty2017ss,kempisty2014aip}, and a stoichiometric (4$\times$4) reconstruction is stable in case of semiinsulating bulk, where stabilization mechanism is based on the charge conservation principle~\cite{strak2024arxiv}.
Since the free energies of two structures are close to each other around the conditions corresponding to the phase boundary between them, it is considered that the mixture of these structures actually emerges. 

In this study, thus, a heterogeneous reconstruction as a mixture of two different reconstructions is investigated. Especially, we focus on a mixture of Ga$_{ad}$(2$\times$2) and \mbox{3Ga-H(2$\times$2)} as a possible example, where surface coverages are 0.25 for Ga$_{ad}$(2$\times$2) and 0.75 for 3Ga-H(2$\times$2). The simplest mixture model is created by arranging small motifs in a regular pattern based on two-dimensional lattice vectors. Here, motif means a template with fixed adatom sites based on a known reconstruction of small surface periodicity. Specifically, the motifs here correspond to Ga$_{ad}$(2$\times$2) and \mbox{3Ga-H(2$\times$2)}, and each defines the number and the site of Ga/H adatom within a (2$\times$2) area. In this study, the simplest such models are considered baseline models, and models with more complex configurations are investigated.

However, the use of a larger lateral cell to represent such mixed structures leads to a combinatorial explosion in the number of candidate structures. It is no longer possible to perform first-principles calculations for all of the enormous number of candidates. One approach to tackle this difficulty is the efficient sampling of relatively stable structures through Bayesian optimization \cite{snoek2012practical,ueno2016combo}, which has been attracting attention as a machine learning application to materials researches \cite{ono2022optimization,hou2019machine,ju2017designing,seko2015prediction}. Recently, a search for the stable configurations of adatoms in the GaN(0001)-(6$\times$6) surface slab model by Bayesian optimization within a certain surface composition has been reported \cite{kusaba2022exploration}. The surface composition corresponds to a surface system consisting of three Ga$_{ad}$(2$\times$2) lateral cells and six 3Ga-H(2$\times$2) lateral cells, which will be adopted as a test system also in this paper. For Bayesian optimization to work properly, the number of candidates should not be too enormous. In that literature, the number of candidates is reduced by imposing reasonable constraints on adatom configurations. In general, however, such reasonable constraints can not be necessarily set for any surface systems. So, how can a widely applicable approach for the study of complex surface structures be developed? In addition, there is no way to confirm that the true optimal structure is contained in the limited number of samples by Bayesian optimization. How can an approach be developed for whole number evaluation? To explore clues to these, this study begins with an analysis of the structures sampled by Bayesian optimization.

\section{SYSTEM DEFINITION}
In our test system, i.e., a mixture of Ga$_{ad}$(2$\times$2) and 3Ga-H(2$\times$2) within GaN(0001)-(6$\times$6) lateral cell, the surface site for Ga adatom is on surface N atom (called as T4 site)\cite{smith1997reconstructions}, and the surface site for H adatom is on surface Ga atom (called as Top site) \cite{fritsch1998ab}. 
If we introduce two types of motifs with the size of (2$\times$2) cell as the minimum unit for adatom arrangement, we can consider the simplest class of the mixtures by placing the motifs within (6$\times$6) cell. Here, one motif consists of an empty Top site and three H adatoms, and the other motif consists of three empty T4 sites and a Ga adatom that bonds with three surface Ga atoms inside the cell, corresponding to the (2$\times$2) reconstructed structures \cite{kusaba2022exploration}.
But in this work, suppose that the two surface reconstructions are mixed with a higher degree of freedom rather than maintaining the (2$\times$2) motifs. For the number of candidates not to be too enormous, two types of constraints are imposed. As the first constraint, all three Ga adatoms are fixed in a certain arrangement. In this circumstance, while the system contains 18 H adatoms, there are 27 sites for H adatoms. Here, the Top sites nearest to the Ga adsorbed sites are excluded, since a Ga adatom form bonds with the three nearest surface Ga. Thus, the number of whole candidates in this test system is up to $_{27}C_{18}$ = 4,686,825. As the second constraint, an empty site and three H adatoms are fixed as a 3Ga-H(2$\times$2) motif. Then, the number of candidates is reduced to $_{23}C_{15}$ = 490,314. These candidates can be considered as a subset within the whole candidates. For further descriptions of our test system, please refer to the literature \cite{kusaba2022exploration}.

\section{DATA SAMPLING}
Bayesian optimization, which was combined with density functional theory (DFT) calculations, sampled relatively stable structures from the subset of candidate structures. Figure 1 shows the sampling histories. A series of sequential sampling was conducted twice (the 1st and 2nd runs hereafter). In each run, only the first two structures were sampled randomly as initial data. Then, Bayesian optimization attempts to sample structures with lower and lower mixing enthalpy. 
Here, the mixing enthalpy is defined as follows,
\begin{eqnarray}
E_{mix} =&& E_{mixture}[\rm{18H3Ga}] \nonumber\\
&& -\frac{2}{3}E_{pure}[\rm{27H}] -\frac{1}{3}\it{E_{pure}}[\rm{9Ga}],
\end{eqnarray}
where $E_{mixture}[\rm{18H3Ga}]$ is the total energy of our test system, $E_{pure}[\rm{27H}]$ and $E_{pure}[\rm{9Ga}]$ are the total energies of the pure systems based on the (2$\times$2) motifs. These total energies are obtained from DFT calculations using GaN(0001)-(6$\times$6) slab models. The details of the calculation as well as optimization setup are in accordance with the literature \cite{kusaba2022exploration}.
With sampling in progress, decreasing trends in mixing enthalpy can be certainly observed. But, more trials were needed for the 2nd run to be sufficiently optimized than for the 1st run. This difference could primarily result from random-like explorations in the early stages. After the declining trend saturated, we continued to optimize as far as our computational resources, and the Bayesian optimizations were finally terminated after 326 and 357 structures were sampled in the 1st and 2nd runs, respectively.
During the optimization, a lot of structures stabler than the best baseline model (dashed line in Fig. 1), which is the most stable among the models created by arranging the motifs regularly, were obtained.

\begin{figure}
\includegraphics[width=8.5cm]{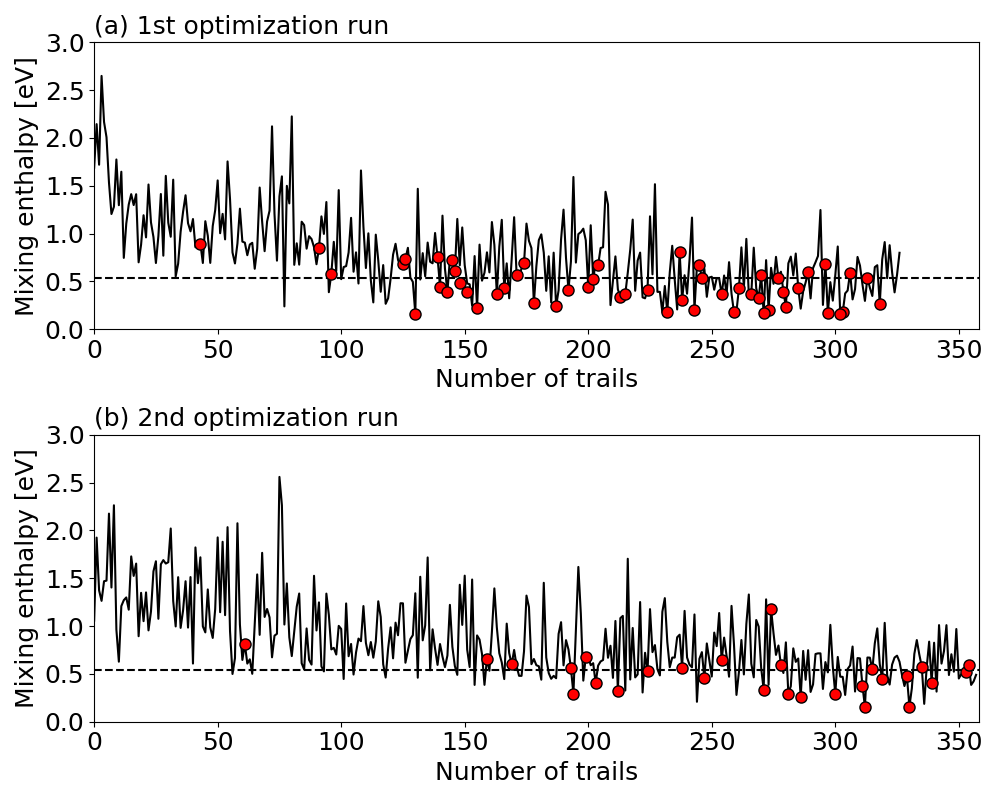}
\caption{History of sampling structures by Bayesian optimization in (a) the 1st run and (b) the 2nd run. Dashed line indicates the mixing enthalpy of a baseline model based on the (2$\times$2) motifs. Red circles indicate the structures satisfied with the local EC rule.}
\end{figure}

\section{ANALYSIS AND DISCUSSIONS}
\subsection{Local electron counting rule}
Electron counting (EC) rule \cite{pashley1989electron} is useful enough to discuss the stability of semiconductor surface. 
While more precise extended electron counting rule (EECR) has been proposed \cite{kempisty2017ss,kempisty2014aip}, it requires DFT extensive calculations and it is not suitable in our large
configuration number system.
Actually, the surface reconstructions Ga$_{ad}$(2$\times$2) and 3Ga-H(2$\times$2) are stable because they satisfy the EC rule. That is, all electrons in Ga dangling bonds that are energetically unfavorable are transferred to the newly emerged Ga–Ga or Ga–H bonds without excess or deficiency of electrons. All candidate structures, which are mixtures of them, also satisfy the EC rule to the extent of (6$\times$6) lateral cell. But this situation can require longer electron transfer, and the satisfaction in such a sense does not necessarily contribute to stabilization. Actually, in the early stages of sampling by Bayesian optimization, unstable structures were observed. Thus, in our (6$\times$6) system, the stricter satisfaction to the extent of (2$\times$2) lateral cell is considered (the local EC rule hereafter). 
That is, using the local EC rule, we then analyze the structures already obtained from Bayesian optimization to investigate the character of the relatively stable mixtures.
Satisfaction of the local EC rule is judged by the following tiling operation.
\begin{figure}
\includegraphics[width=8.5cm]{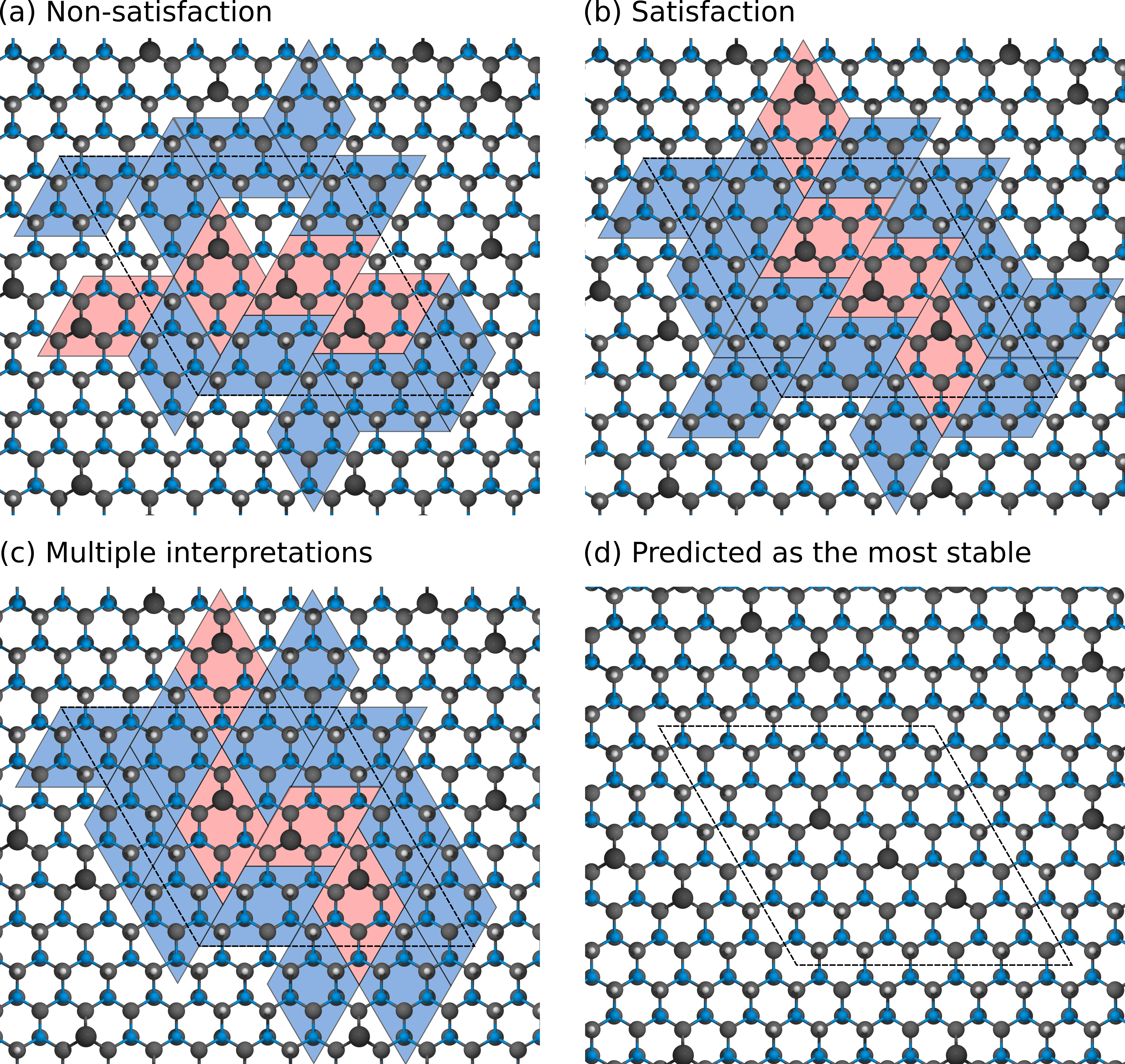}
\caption{Examples of tiling operation to judge satisfaction of the local EC rule. Tile A is shown in red shading, tile B in blue shading. Dashed line indicates (6$\times$6) supercell. The structures (i.e., configurations of adatoms) in (b) and (c) are same. Gray and blue balls indicate the Ga and N atoms of substrate, respectively; small white ball indicates H adatom; and larger darker black ball indicates Ga adatom.}
\end{figure}
\begin{itemize}
      \item In a (6$\times$6) structure under consideration, lay out three (2$\times$2)-size tiles A and six (2$\times$2)-size tiles B with no overlaps or gaps.
      \item Within tile A, one Ga adatom with three nearest-neighbor surface Ga and no H adatoms must be included, which corresponds to Ga$_{ad}$(2$\times$2) motif.
      \item Within tile B, three H adatoms and one empty site must be included, which corresponds to \mbox{3Ga-H}(2$\times$2) motif. But, the internal configuration of H adatoms within each tile is not restricted.
\end{itemize}
Figure 2 shows examples of the results of tiling operation. The structure in Fig. 2(a) does not satisfy the local EC rule, because there are gaps between tiles. Correspondingly, only five tiles B are laid out. The structure in Fig. 2(b) satisfies the local EC rule because the tiles can be laid out without overlaps or gaps. 
Note that the valid tiling layout is still disorderly and complex compared with the baseline model created by arranging the motifs regularly.
The tiling operation for all structures sampled by Bayesian optimization was performed manually like some sort of jigsaw puzzle. \mbox{Figure 1} shows the structures that satisfy the local EC rule with red circles. It can be seen that most structures that satisfy the local EC rule are relatively stable, among all the structures sampled. However, the stability of those structures is fluctuating. What differences between the structures that equally satisfy the local EC rule can be attributed to this fluctuation? Furthermore, the stable structures in spite of not satisfying the local EC rule can also be observed. This means that the local EC rule defined above is insufficient to fully explain the stability of the mixture. 

\begin{figure}
\includegraphics[width=8cm]{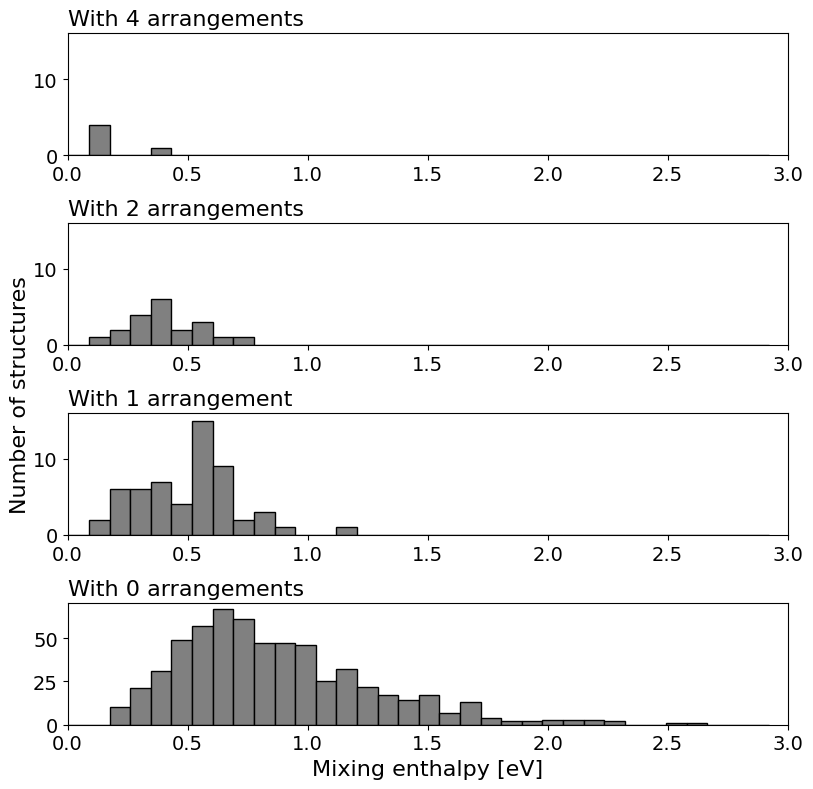}
\caption{Relation between the number of arrangements and stability. After all sampled structures are classified by the number of arrangements, the mixing enthalpy distributions by the classes are displayed in histograms.}
\end{figure}

Next, the concept of the number of arrangements for the local EC rule is introduced. For a given structure, tiling layout is not necessarily uniquely determined. While there is a structure that does not have even a single valid tiling layout, there is also a structure that has more than one of layouts. For example, Fig. 2(c) shows the same structure as Fig. 2(b), but the tiling layouts shown in (b) and (c) are different. The number of tiling layout a structure has is referred to here as the number of arrangements for the structure. \mbox{Figure 3} illustrates how the number of arrangements relates to the stability of the structure. Among the structures with the same number of arrangements, there are variations in stability. But, as the number of arrangements increases, the maximum mixing enthalpy decreases. In other words, structures with higher number of arrangements can not be unstable. Therefore, it can be said that the tiling operation has a potential as an effective method for generating only stable structures.
On the other hand, the minimum mixing enthalpies for each arrangement number are comparable to each other. In other words, structures with even lower numbers of arrangements can be stable. Thus, even introducing the concept of arrangement number, it was not still achieved to fully explain the stability of the mixture. However, the facts in the above analysis inspire the following hypothesis: the local EC rule fails to evaluate the inter-tile interactions depending on the intra-tile adsorption configurations. This hypothesis is consistent with the intuition that another arrangement would allow the inter-tile interactions in one arrangement to be evaluated in the framework of the local EC rule. Also, the structures could be possible that sufficiently relax the inter-tile interactions, even if they do not strictly satisfy the local EC rule.

\subsection{Data-driven Ising model}
An Ising model \cite{brush1967history,newell1953theory} is invoked to overcome the problem hypothesized above. Here, interactions between adatoms are evaluated beyond the tile boundaries in an omnidirectional, but short-range, manner. The Ising model is expected to explain the stability of the mixture, at least qualitatively, by having the following mechanisms.
\begin{itemize}
      \item Empty sites repel each other, and H adatom and empty site attract each other.
      \item Ga adatom repels with H adatom and also attracts empty site.
\end{itemize}
The former mechanism arranges the empty sites uniformly and shortens the electron transport distances from Ga dangling bond to Ga-H bonds. This is a generalization of the stabilization mechanism of the 3Ga-H(2$\times$2) reconstruction. The latter mechanism shortens the electron transport distances from Ga dangling bond to Ga-Ga bonds. This is a generalization of the stabilization mechanism of the Ga$_{ad}$(2$\times$2) reconstruction. These mechanisms are implemented in an Ising model as follows.
\begin{eqnarray}
\hat{E}_{mix} = -\sum_{i<j} J\sigma_i\sigma_j -\sum_i h\sigma_i +c,
\end{eqnarray}
where $\hat{E}_{mix}$ is mixing enthalpy; $\sigma_i$ is a variable that takes 1 when H adatom exists at a surface site $i$ and $-$1 when the site $i$ is empty; $J$, $h$ and $c$ are parameters of this model. 
The sum of the 1st term is performed for all nearest-neighbor site pairs. The 1st term represents the 1st mechanism mentioned above if the parameter $J$ takes a negative value. This term also has the effect of repelling H adatoms from each other, but this is a coinside to the mechanism of distributing empty sites uniformly and is not considered harmful. 
The sum of the 2nd term is performed for all the 2nd nearest-neighbor sites of Ga adatom. 
Note that the 1st nearest-neighbor sites are not included to the possible sites for H adsorption in our system.
The 2nd term represents the 2nd mechanism mentioned above if the parameter $h$ takes a negative value. The parameter $c$ is the bias term.
Since Ga adsorption sites are fixed in our current search space, Ga-Ga interaction is included in the bias term while Ga-H interaction is included in the 2nd term.

\begin{figure}
\includegraphics[width=6cm]{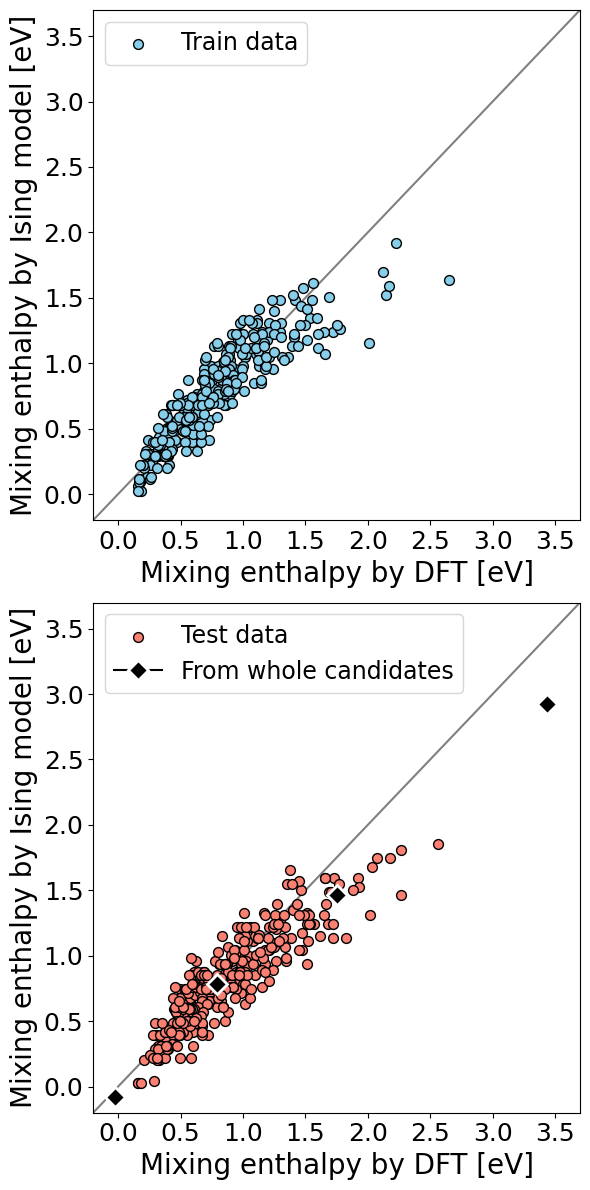}
\caption{Comparison of mixing enthalpies estimated by Ising model and calculated by DFT. Structures as train data (blue circles) and test data (red circles) are sampled by Bayesian optimization from the subset of candidates up to 490,314 structures. To validate the prediction ability of the trained model to a wider range of candidate structures, four structures (black squares) were sampled from the whole search space with 4,686,825 structures.}
\end{figure}

Parameter determination for the Ising model is performed using data of the sampled structures by Bayesian optimization. That is, the sampled structures are expressed in $\sigma_i$ and input into the Ising model; the parameters are determined to minimize the sum of square error, $SSE \equiv \sum_i^N (\hat{E}_{mix,i}-E_{mix,i})^2$,where $i$ is the index of structure and $N$ is the number of data, between the mixing enthalpies estimated by the Ising model and those calculated by DFT. The data used to determine the parameters (i.e., train data) were all data from the 1st run. 
The resulting parameters are as follows: $J=-0.010932$~[eV]
, $h=-0.097453$~[eV], $c=0.818264$~[eV].
Thus, in a sense per pairwise, the interaction between Ga and H adatoms is more dominant than the interaction between H adatoms.
A comparison between the value from the Ising model with the parameters determined in this way and the value from DFT calculations is shown in Fig. 4 (blue circles). A clear linear relation can be seen in the plot, and root mean squared error, $RMSE$, is 0.18~eV, where $RMSE \equiv \sqrt{SSE/N}$.
Thus, it can be said that the Ising model sufficiently quantitatively explain the stability of the mixture
although it underestimates in high mixing enthalpy range, where complex interaction might go beyond the superposition of the nearest-neighbor pairwise interactions. 
Next, to check that no overfitting to train data has occurred, all data from the 2nd run (i.e., test data) were input to the model with the parameters determined by the 1st run data. The result is shown in Fig. 4 (red circles), and $RMSE$ is 0.19~eV, which is comparable to the value for train data. Thus, it was confirmed that our Ising model with the obtained parameters has the ability to predict mixing enthalpy, at least to the extent of the subset of candidate structures, which consists of up to 490,314 structures.

Next, whether the model and the parameters have the prediction ability for the whole candidate structures, which are up to 4,686,825 structures, needs to be checked. It is expected to be true because introducing the above mentioned constraint, which is for an empty site and three H adatoms, does not cause loss of generality in the discussion of the two stabilization mechanisms. \mbox{Figure 5} shows the histogram of mixing enthalpies obtained by inputting all 4,686,825 candidate structures into the model. In other words, it is an approximate full picture of the whole candidate structures that could not be revealed only by DFT calculations and Bayesian optimization sampling. It can be seen that the structures more stable than the baseline model (see the literature \cite{kusaba2022exploration}) are a very small percentage of the whole candidate structures. That is, the efficiency of Bayesian optimization sampling was once again confirmed. Next, some structures were picked up from the whole candidates, including the structure predicted to be the most stable (shown in Fig. 2(d)) based on the estimated energy structure of Fig. 5, and the values of mixing energetalpy were confirmed by DFT calculations. The result is shown in \mbox{Fig. 4} (black squares), the mixing enthalpy of the structure predicted as the most stable is $-$0.08~eV for the model and $-$0.02~eV for DFT. The plots of the picked structures showed a linear relation as well as samples from the subset of candidate structures, which confirmed the prediction ability for the whole set of candidate structures. Although there are of course some errors, the predictions are quantitative enough for purposes such as finding the most stable structure or giving an overview of the energy structure.

\begin{figure}
\includegraphics[width=7.3cm]{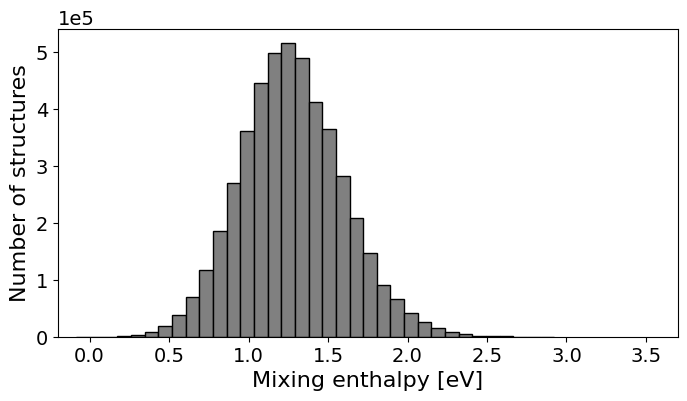}
\caption{Estimated energy structure of the mixture system consisting of 4,686,825 states (i.e., configurations of adatoms).}
\end{figure}

\section{CONCLUSIONS}
In summary, by generalizing the concept of the local EC rule, an Ising model has been proposed that can quantitatively predict the mixing enthalpy of the two different surface reconstructions. Utilizing this Ising model, it is possible to exhaustively evaluate an enormous number of all candidate structures, which was not possible with conventional study approach by DFT calculations. In other words, we have discovered an approach for revealing a realistic surface structure by efficient data collection using Bayesian optimization and a data-driven Ising model.
This approach would enable distribution function calculations for mixtures of surface reconstructions. Also, as a possible application, it can provide a realistic reaction field for surface reaction research.
Furthermore, it is expected that our scheme can be applied to any semiconductor surface for which the EC rule is valid, and that it will contribute significantly to the advancement of surface science.

\begin{acknowledgments}
The author A.K. wishes to acknowledge Professor T. Kuboyama for his valuable advice.
This work was partially supported by JSPS KAKENHI (grant numbers JP20K15181, JP23H03461); Collaborative Research Program of Research Institute for Applied Mechanics, Kyushu University; and Diversity and Super Global Training Program for Female and Young Faculty (SENTAN-Q), Kyushu University. 
The computation was carried out using the computer resource offered under the category of General Projects by Research Institute for Information Technology, Kyushu University; and under computational allocation No. G91-1448 by Interdisciplinary Centre for Mathematical and Computational Modelling, University of Warsaw.
\end{acknowledgments}



%

%

\end{document}